# Astrophysical sense of the gravitational waves discovery

V. M. Lipunov

Lomonosov Moscow State University , Sternberg Astronomical Institute, Physics Department
Universitetskii prosp. 13, 119991 Moscow, Russian Federation; Lomonosov Moscow State University, Faculty of Physics, Leninskie gory 1, str. 2, 119991 Moscow, Russian Federation
E-mail: lipunov@sai.msu.ru

*Abstract*. **The discovery of gravitational waves by the international collaboration LIGO on the one hand is a triumphant confirmation of the general theory of relativity, and on the other confirms the general fundamental ideas on the nuclear evolution of baryon matter in the Universe concentrated in binary stars. LIGO/Virgo may turn out to be the first experiment in the history of physics to detect two physical entities, gravitational waves and black holes.**

## 1. Introduction

On 14 September 2015, the upgraded interferometer system LIGO (Laser Interferometer Gravitational Wave Observatory) for the first time detected the gravitational waves generated by merging binary black holes of roughly equal masses at a distance of more than one billion light years from Earth [1].

On the one hand, this discovery was anticipated by Lipunov et al. in 1997 within the modern binary stellar evolution theory [2-4]. On the other hand, for the first time in the history of astronomy, the discovery established the true interworkings of a new information channel - gravitational waves - with electromagnetic terrestrial and space observatories [5]. Owing to the pioneering work of Soviet astrophysicists on binary star population synthesis (the Scenario Machine [6]) and the development of the MASTER global robotic telescope network, we managed to take an active part in creating a new science: gravitational wave astronomy.[1] The particular importance of the discovery is that two objects predicted by general relativity were concurrently discovered: gravitational waves and black holes.

Originally, an experiment to detect gravitational waves was discussed by Gertzenshtein and Pustovoit [7]. Later, it was further developed by Braginsky [8-10]. Moreover, Russian scientists were directly involved in the experiment itself [5, 8, 11, 12]. Thus, Soviet and Russian scientists notably contributed to the discovery of gravitational waves.

In 1964, Zeldovich noted that observations of binary relativistic stars are of particular interest [13]. The existence of relativistic stars in binary systems became evident already in the first evolutionary scenarios of massive binary stars [14-16]. Relativistic stars --- neutron stars (NSs) and black holes (BHs) - are formed from massive stars (more than $10M$ ) able to produce nuclei with masses exceeding the Chandrasekhar limit for a white dwarf (~$1.5M_\odot$) and the Oppenheimer-Volkoff limit for a neutron star (~$2.5M_\odot$) at the end of the thermonuclear evolution.

---

[1] The author of this paper was lucky to contribute to the discovery of gravitational waves on two occasions. First, as a theorist who proposed the Scenario Machine. Second, since the second half of 2015, as an experimentalist who is taking charge of the MASTER global robotic telescope network.

Such processes can also occur in so-called low-mass systems. But the formation of binary relativistic systems - binary neutron stars (NS + NS) or black holes (BH+BH) as well as mixed pairs (BH + NS) - is possible only as a result of the evolution of two massive stars, either of which can form a relativistic star. Already in the 20th century, when a binary neutron star was detected in our Galaxy [17] (1993 Nobel prize), it became clear that binary relativistic stars can be quite powerful sources of gravitational waves. The study of such systems explicitly showed that general relativity is correct [up to $\sim (v/c)^5$ ], in full accordance with Einstein's formula [18] (1916) for the gravitational wave power:

$$L = \frac{32}{5} \frac{G^4}{c^5} \frac{M_1^2 M_2^2 (M_1 + M_2)}{A^5} .$$

Here, G is the gravitational constant, $M_1$ and $M_2$ are the masses of binary system components, A is the distance between them (with the orbits assumed to be circular), and c is the speed of light in a vacuum.

The merging of two stars is the most powerful macroscopic process in the Universe. Indeed, we consider two massive sources colliding with each other at the maximum velocity c. The power of the process is $L \approx E/t_{min}$, where $E \approx Mc^2$ and the minimum time is $t_{min} = R_{min}/c$, while the minimum radius $R_{min}$ of any body is the gravitational radius $R_{min} = R_g = 2GM/c^2$. We can easily show that the maximum power (or, as astrophysicists say, luminosity) is $L_{max} = c^5/G \approx 4.5 \cdot 10^{59}$ erg s$^{-1}$

Einstein called this power natural luminosity.

It is remarkable that if the Planck energy $E_{Pl}=(hc^5/G)^{1/2}=1.22 \cdot 10^{28}$eV is divided by the Plank time $t_{Pl}=(hG/c^5)^{1/2}=5.39116 \cdot 10^{-44}$ s,
the Planck constant drops out from the luminosity formula, and we find the natural luminosity again [19]:

$$L_{Pl} = \frac{E_{Pl}}{t_{Pl}} = \frac{c^5}{G} = 4.5 \times 10^{59} \text{ erg s}^{-1}$$

This is the power at which the Universe was being born.

It is no accident that some of the most powerful electromagnetic bursts observed in the Universe - short gamma-ray bursts - are related to colliding neutron stars with radii close to the minimum one.

It follows that natural luminosity will play an important role even in a future theory of quantum gravity.[2] The most powerful macroscopic processes in the Universe are colliding relativistic stars. It is the observation of the binary radio pulsar PSR 1913+16 that demonstrated that the merging time is less than the Hubble time (*1/H $\approx$ 14* billion years, where *H $\approx$70 km s$^{-1}$ Mpc$^{-1}$* is the Hubble parameter). It formed the basis for the first experimental estimates of the coalescence rate of neutron stars in the Universe and the probability of detecting this process[12]. Later, it became clear that the merging of neutron stars may have already been detected as short gamma-ray bursts.

---

[2] Of course, luminosity is not Lorentz invariant, because both the energy and the energy release time change when moving relative to the observer. But this is not important from the standpoint of astrophysics, because there are no relativistic macro sources moving towards us in the currently expanding Universe.

By the early 1980s, it became clear that there are specific macroscopic reactions (M-reactions) of the `elementary particles' of the Universe - neutron stars and black holes - with the maximum possible power [3] ~ $c^5/G$ ~ $10^{59}$ erg s$^{-1}$ [20]:

NS + NS → BH + GWB + EMB
or
NS + NS → NS + GWB + EMB
if the sum of NS masses is below the Oppenheimer-Volkoff limit ;
NS + BH → BH + GWB + EMB ;
BH + BH → BH + GWB .

Here, GWB and EMB are gravitational and electromagnetic wave bursts. In the first process, there are two possible outcomes depending on the upper mass limit of the neutron star (the Oppenheimer-Volkoff limit), which is not exactly fixed yet.

By the early 1980s, the `cross section' (or, say, the probability) of such reactions was unknown. In particular, it was not clear which of the processes are more common in the Universe. Understandably, the LIGO experiment results depend on the maximum-sensitivity frequency.

The frequency of gravitational waves emitted by a binary system is determined by the rotation frequency: more precisely, it equals twice the orbiting frequency $\Omega = 2\pi/P$. From the third Kepler law, we find that $P^2/A^3$ ~ $(M_1 + M_2)^{-1}$.

At the same time, the minimum distance is proportional to the mass: A ~ $M_1 + M_2$ (the Schwarzschild radius). It follows that the maximum frequency of a binary system at the collision moment - the merging of components - is *v=2/P~1/($M_1$ + $M_2$)*. Because the black hole mass can be a few dozen times larger, their frequency *n* is one order of magnitude less than the orbiting frequency of neutron stars.

Estimates have shown that the first events should have frequencies in the range 100-200 Hz (!) rather than the 1000 Hz typical for neutron stars. In other words, the gravitational wave detector should have a wide tuning range! But this is a question of time and money. For example, solid-state detectors and even some interferometers were originally tuned to the frequency of the order of 1kHz.

## 2. Scenario Machine

In the early 1980s, a new method was proposed to study the late stages of stellar evolution, the population synthesis of binary stars including the formation and evolution of relativistic stars: neutron stars and black holes [21, 22]. In the first papers, in particular, the formation of relativistic binary systems with black holes was demonstrated. The method has allowed the first calculation of the expected coalescence rate for binary NSs, normalized to a constant star formation rate typical for our Galaxy with the mass $10^{11}M_\odot$ [20]. It also became possible to compute the amplitude and the continuous spectrum of the gravitational wave background generated by binary stars [23]. Similar results were independently obtained in [24], where black hole merging processes were also considered.

---

[3] For example, the luminosity of quasars, considered in the 20th century as most powerful objects in the Universe, is 10 orders of magnitude less.

But it was the Scenario Machine that in 1997 for the first time allowed Lipunov et al. [2-4] to show that merging binary relativistic systems including black holes, BH + BH and BH + NS, should be the first events on LIGO-type detectors (Fig. 1).

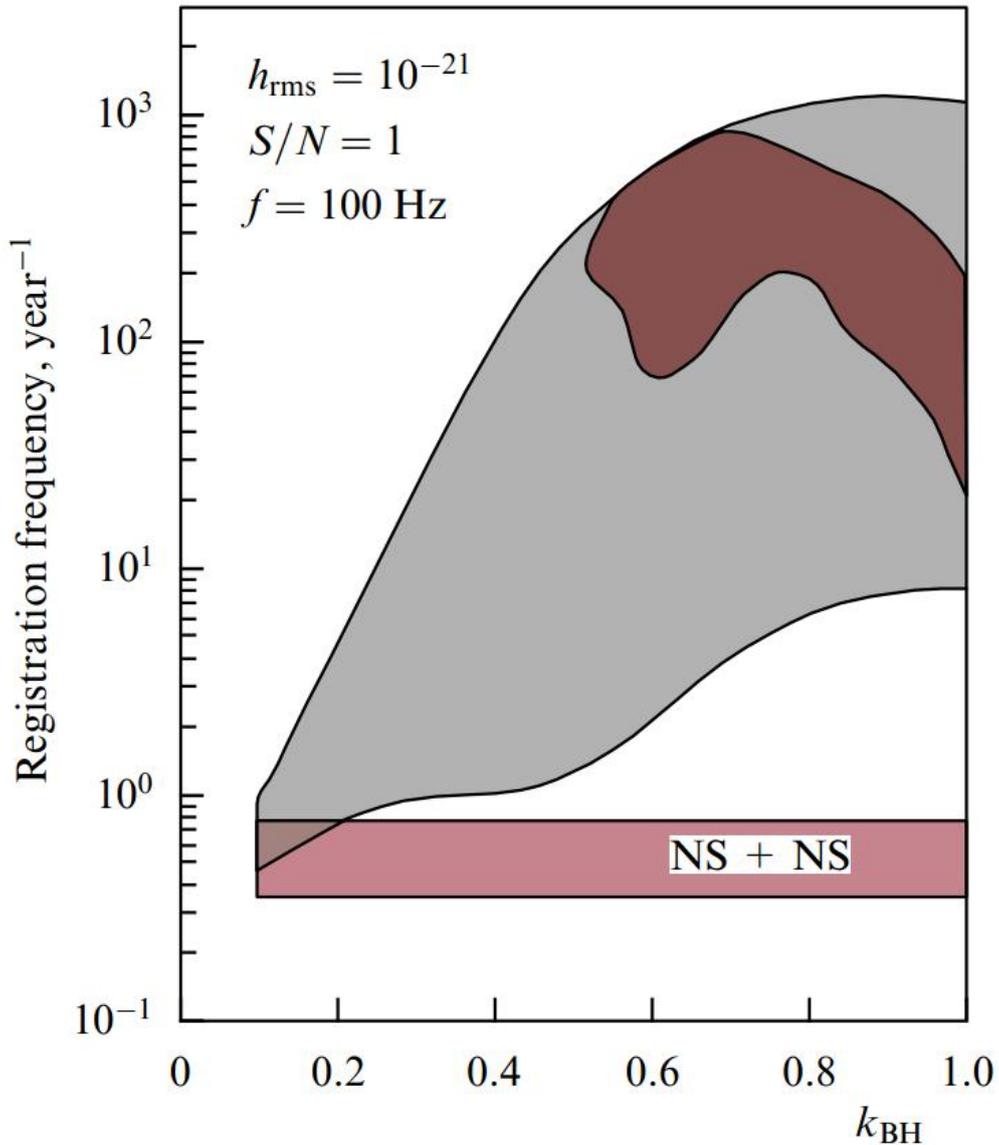

*Figure 1. `Loch Ness monster' (a dinosaur head). An expected registration rate (by a detector with the sensitivity $h_{rms} = 10^{-21}$ at the frequency $f = 100$ Hz with the signal-to-noise ratio S/N =1) of the gravitational wave bursts generated by NS and BH merging events depending on the yet unknown parameter $k_{BH}$ - the portion of stellar matter escaping towards a black hole at the moment of the back hole formation. Grey area shows the black hole merging rate on LIGO-like detectors for all possible parameters of the binary evolution scenario with a weak stellar wind. Looking similar to the head of a prehistoric monster, the brown area shows the region of possible registration rate found from the modern binary stellar evolution theory. This region is quite large because of many unknown parameters. However, the region lies everywhere substantially above the region of the signal registration frequency from merging neutron stars (the horizontal line marked by NS + NS). The diagram shows that merging black holes must be detected first [2].*

The subject of gravitational waves is close to studies of gamma-ray bursts, in particular, of so-called short gamma-ray bursts, which are related to mergings of neutron stars and mixed systems. However, we should not expect gamma-ray bursts to be directly identified with neutron-star mergings: gamma-ray emission is strongly anisotropic and concentrated in a small solid angle of the order of a few degrees; therefore, most gamma-ray bursts do not hit Earth. For example, the probability of simultaneous detection of gravitational-wave and gamma-ray bursts (GRBs) is 1/1000. This estimate is important in Section 7, where we analyze reports on a possible detection of a gamma-ray burst in the event GW150914 by the Fermi observatory.

However, it can be argued that quasi-isotropic electromagnetic radiation must form during and before the collision moment, and mergers involving neutron stars must be followed by an afterglow.

In 1984, Blinnikov et al. [25] were the first to show that neutron star mergings can be supplemented by a powerful electromagnetic burst. In 1996, Lipunov and Panchenko showed [26] that powerful nonthermal pulsar-like radiation (in terms of the formation mechanism rather than periodicity) is possible during and before neutron star mergings. Faded pulsars can also flash for a moment. This is not due to spinning rotation, as usual, but due to orbital motion, which attains a kilohertz frequency at the late stages of merging. In this case, appropriate conditions are created to generate regions near the neutron star surface where electric and magnetic fields are parallel, while magnetic field lines are not closed.

Using the standard reasoning based on the Pointing vector of the electromagnetic energy flux, we can find that by the time of collision, the nonthermal luminosity (including X-ray and radio emission) can be a few million times larger than the luminosity of the known radio pulsar in the Crab Nebula for equal values of the magnetic field.

In the case of a nonstandard magnetic field (H ~ $10^{13}$ -$10^{14}$ G), the luminosity can increase 10,000-fold, such that a pulsar can be visible from distances up to 100 Mpc. Incidentally, this exceeds the sensitivity horizon of the upgraded LIGO interferometer [1] with respect to neutron star mergings.

After the merging, part of the radioactive matter in which heavy elements are being fused can be ejected; a so-called kilonova may occur in a time period from several hours to several days [27]. Another process occurring during the merging is the temporary formation of a spinar, which is a rapidly rotating self-gravitating object [28]. We stress once again that we here do not consider gamma-ray bursts with electromagnetic radiation concentrated in a narrow jet [29, 30], because the probability of its detection in the first successful observations of gravitational waves is very small.

After mid-2015, the Russian system of wide-field robotic telescopes, MASTER, joined the EM follow-up program of the gravitational wave LIGO-Virgo experiments [5]. We say a few words on how the MASTER system originated. In 2003, supported by a private sponsor, we started to develop robotic observations of astronomical explosive objects [31], primarily, early observations of proper optical emission of gamma-ray bursts. To date, with support from the developing program of Moscow State University and the joint stock company OPTIKA Moscow Consortium, we have developed a global network of identical wide-angle telescopes localized in the Northern and Southern hemispheres [11, 32-34] (Fig. 2).

On 16 September 2015 at 05:39:58 UT, we obtained the error probability matrix of the first gravitational wave alert ALIGO trigger G184098 [35]. On the following night, we started to survey candidate regions by all MASTER network telescopes (Figure 2). We examined a region of the sky with an approximate area of 5000 $deg^2$ with various limits up to 20 optical magnitude (Figure 3,4). Those results were briefly described in the joint paper of the LIGO/Virgo collaboration EM follow-up groups [5]. More details can be found in the MASTER collaboration paper [12].

## 3. MASTER global robotic telescope network

We first state our conventions on terminology. Participants in regular international conferences on robotic observatories and telescopes defined a robotic telescope as follows (Robotic Autonomous Observatories Workshop 2009): a robotic telescope is able to make multi-day observations without human input, automatically receiving and processing images, saving new data in its own database and sending updates by emails and telegrams. Of course, such a telescope can be maintained remotely via the Internet. However, manual operation is rare and, in general, impairs the effectiveness. In some cases, the MASTER robot-software sends a scientific telegram, and when the recipient is another robot, publication is automatic. The MASTER telescopes already operate in detecting potentially hazardous asteroids moving with a large angular velocity and in observing gamma-ray bursts when events last a few dozen seconds.

The idea of creating the MASTER network [34] was to install completely identical MASTER II robotic telescopes in both the Northern and Southern hemispheres on eastern and western longitudes [11]. Each MASTER II telescope consists of two wide-field 400 mm telescopes with a field angle of 4 $deg^2$ installed on a single super-rapid mount (with the aiming speed 20° -30° per s) supplemented by a third axis, which brings the telescope tubes together and drives them apart. In the parallel position, this device allows obtaining synchronous images of rapidly changing objects using various filters or polarization directions. MASTER II is the only wide-field color telescope in world that can measure the polarization of rapidly changing objects. The parallel tubes are typically used in `alert' observations of gamma-ray bursts, generally, in two polarizations. Presently, the MASTER network is leading the early optical observations of gamma-ray bursts [36].

Recently, MASTER telescopes were able to detect the earliest polarization of the optical radiation of gamma-ray bursts [37]. However, most of the time, the MASTER telescopes are involved in the regular sky survey aimed at detecting new objects unreported in world catalogues and the MASTER database itself.

The main advantage of the MASTER network is a unique mathematical software that allows automatically detecting about 10 different types of astrophysical transients. Despite their small size, the MASTER telescopes equally compete with the largest wide-field telescopes in the world [Pan-STARRS (Panoramic Survey Telescope and Rapid Response System), Catalina, iPTF (intermediate Palomar Transient Factory)] detecting astrophysical explosions in the Universe.

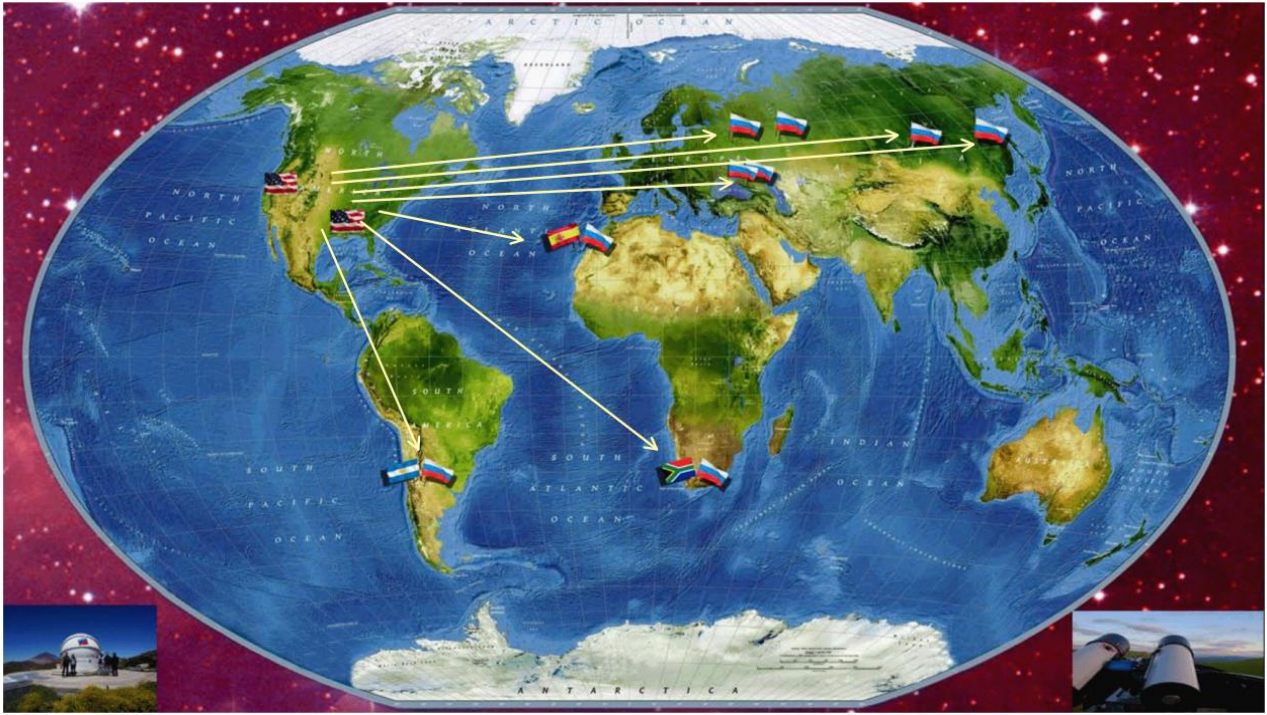

*Figure 2. Locations of the MASTER global network robotic telescopes and the gravitational wave antennas of the American LIGO interferometer. The MASTER telescopes are located (from east to west): near Blagoveshchensk, in the Tunka valley (Tunka Astrophysical Center, Lake Baikal), near Ekaterinburg, near Kislovodsk, in the Crimea, in South Africa, in the Canary Islands, and in Argentina. The flag near Moscow marks the place of the first test MASTER I telescope (MASTER II prototype, now off-line) built in the Domodedovo district. All the MASTER telescopes operate automatically. Having received a signal, they target any available point in the sky in less than 20 s. The MASTER system does the most rapid optical survey of the sky to 19-20 optical magnitude, 64 deg$^2$ per minute.*

Because of the geographic location, the MASTER net-work is a unique search system of wide-field telescopes distributed around the globe. These advantages of the MASTER network were beautifully manifested in optical observations of the first LIGO event on 14 September 2015, significantly contributing to the survey of probable regions of the gravitational wave source (localization) [5].

The MASTER global network currently includes five observatories in and three outside Russia: MASTER-Amur is located near the city of Blagoveshchensk, hosted by Blagovenshchensk Pedagogical State University, MASTER-Tunka is in the Tunka Astrophysical Center of the Applied Physics Institute, Irkutsk State University, MASTER-Ural is in the Kourovka observatory of the Ural Federal University, MASTER-Kislovodsk is located near the city of Kislovodsk at the high-altitude solar station of (the main) Pulkovo observatory, Russian Academy of Sciences, MASTER-Crimea is in Nauchny village at historical Astronomical Station of Moscow State University, MASTER-SAAO is in the South African observatory (RSA, Sutherland), MASTER-IAC is part of the Teide Observatory operated by the Canary Islands Institute of Astrophysics on Tenerife island (Spain), and MASTER-OAFA is in the National University Observatory, San Juan, Andes, Argentina. All observatories are equipped with robotic super-wide field cameras (16 ×24 deg$^2$ with a limit of 11 optical magnitude per second and 13.5 optical magnitude when summing frames).

The super-wide field cameras do nonstop (quick-record mode) sky survey. With the overall area of 5000 deg$^2$, the probability of detecting a gamma-ray burst within the field of view is equal to 1/8, which allows having images of squared errors of bursts synchronously with the initial instant or even preceding ones. In fact, this is the only way to observe proper optical radiation of short gamma-ray bursts whose short duration (less than 4 s) makes alert observations synchronized with gamma-ray observations, impossible even using the super-rapid MASTER network mounts.

During the last three years, the MASTER network has detected more than 1300 exploding objects in the sky. Among them are the prompt optical radiation of gamma-ray bursts (most powerful EM event in the Universe), supernova explosions, including those of type Ia used to test the properties of the dark energy, novae, and dwarf nova stars, quasar explosions and active galactic nuclei (physical plasma in the gravitation field of supermassive black holes), potentially hazardous asteroids and comets, optical transients of a yet unknown nature such as MASTER OT J095310.04+335352.8 (eclipse binary star with an unusually long variability period of about 69 years), or anomalous bright red novae (MASTER OT J0042007.99+405501.1/ M31LRN 2015) (LRN - Luminous Red Nova in Galaxy M31) resulting from the collision of ordinary stars.

## 4. Observation of the gravitational wave event GW150914

After receiving an alert on 16 September 2015, according to weather and nighttime parameters, the MASTER-AMUR, MASTER-Tunka, MASTER-SAAO, and MASTER-IAC telescopes started to examine the areas of localization of the gravitational wave event GW150914 (Figure 1). In just a week, a region of about 5000 deg$^2$ was covered three times. As it turned out later, a part of the region with an area of 560 deg$^2$ was the most probable localization of the gravitational wave event in the southern sky (Figures 3,4). We discovered eight optical transients, three of which were inside or near the GWE final localization region. Later, it turned out that during the standard survey, the day before receiving the first GWE coordinates, we had surveyed the corresponding region with an area of 16 deg$^2$.

The analysis of optical bursts (transients) detected by the MASTER robotic telescopes showed that two of them are dwarf nova stars in our Galaxy. Another object, MASTER OT J040938.68-541316.9, is most probably a type-Ia super-nova. It was discovered shortly before the luminosity reached the maximum. It requires 8-10 days for a Ia supernova to reach maximum luminosity, while this supernova was discovered two days after the gravitational wave trigger. Thus, we can make a reliable conclusion that there is no causal relation between the gravitational wave pulse and the supernova explosion.

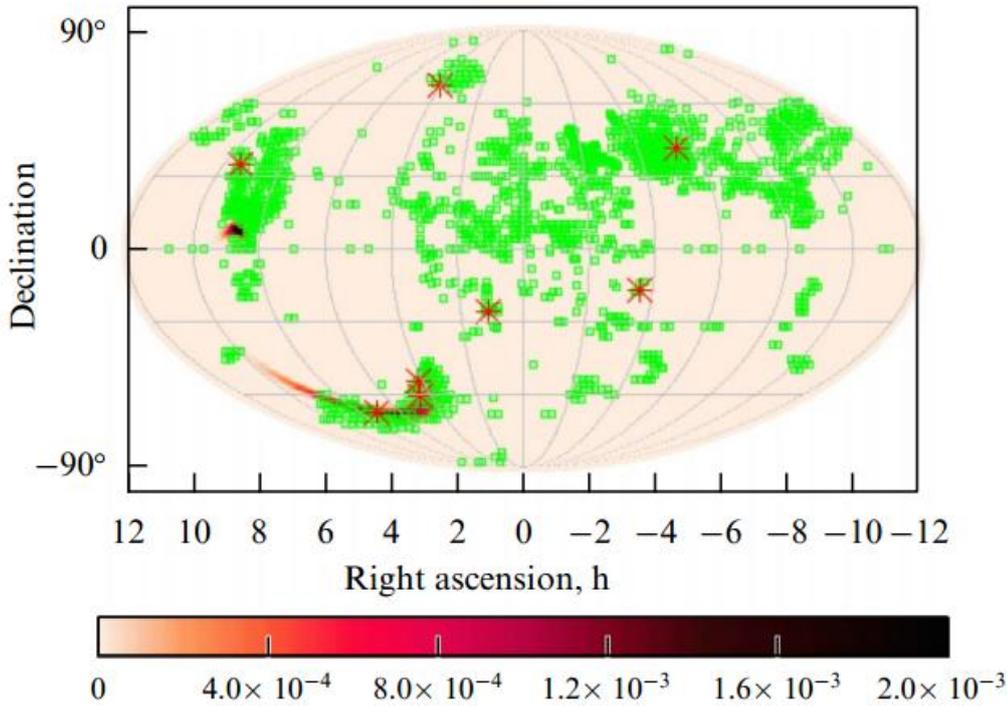

*Figure 3. Sky survey after (and before: unintended filming on 15 September) the alert GW150914/G184098 was received by the MASTER network (green squares). The GRE localization probability is shown in orange. Transients detected by the MASTER network during the survey are marked by red stars*

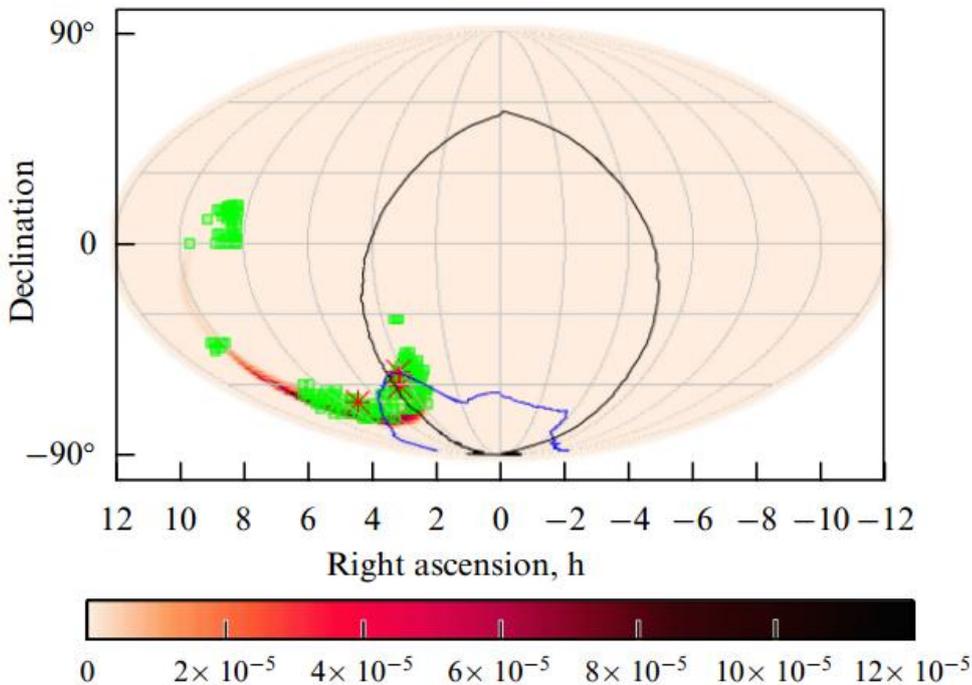

*Figure 4. Final probability distribution of the gravitational wave event LIGO GW150914/G184098 [1]. MASTER scopes of view (in green), three transients detected by them (in red). One-sigma square error of the probable event etected by the Fermi space observatory is shown in blue [12, 39]. The black egg-shaped region is Earth's shadow in observa-tions by the Fermi observatory.*

As noted in [1], the Global Russian MASTER robotic telescope network was instrumental in optical observations of square errors of the first gravitational wave burst in history.

## 5. Why were merging black holes discovered first?

Paper [3] published in 1997 was entitled ``First LIGO events: binary black holes mergings." The gravitational wave burst detected on 14 September 2015 resulted from binary black holes merging [1, 40], which is consistent with predictions of the population synthesis analyzed by the Scenario Machine [2-4]. This discovery confirms the correctness of our ideas about the evolution of binary stars.

Lipunov et al. [2] showed that regardless of the particular evolution scenario and parameters, the first events on LIGO-type interferometers must be black-hole mergings, which is most clearly shown in Fig. 1 [2]. In what follows, we consider the method of obtaining this result in more detail. Moreover, there have been papers claiming that black holes do not merge at all, and cannot therefore be the aim of the first gravitational wave experiments with LIGO-type interferometers.

Indeed, the calculation of the rate of events with black hole mergings is a complicated problem. Simple analytic estimates based on our ideas of nuclear stellar formation face a huge uncertainty related to the initial condition multivariance and the complex structure of evolutionary tracks in binary systems (for more discussion, see [41]).

The special method of population synthesis was proposed by Kornilov and Lipunov in order to analyze various scenarios of binary system evolution and calculate possible parameters of the final products of the evolution: the numerical study of a large number of binary system tracks by the Monte Carlo method (Scenario Machine) [21, 22].

The first calculations by the Scenario Machine immediately gave the statistical properties of various types of massive binary systems, including those at the final stages of stellar evolution resulting in double relativistic systems that are potential sources of gravitational wave pulses during the merging moment. The calculations done in 1987 using the advanced Scenario Machine determined the rate of neutron-star merging in the Galaxy with a particular star formation rate (the Salpeter function) [20, 23].

In 1993, the first calculations of black hole mergings were carried out. It was shown that the black hole merging rate can be comparable to that of neutron stars [24]. However, the evolution of binary stars has a huge number of ill-defined parameters, which made it impossible to find how frequent events could be on LIGO-Virgo-type gravitational wave detectors. The most successful attempt was undertaken using the Scenario Machine by Lipunov et al. [2, 6].

We emphasize that unlike other codes of population synthesis, the Scenario Machine is aimed at comparing the results of numerical studies with all possible observational data on the relativistic stages of binary stars: radio pulsars in binary systems with different types of components, X-ray pulsars, black hole candidates, millisecond pulsars, etc. This allowed choosing the optimal parameters of the stellar evolution in such a way that the observable distribution of neutron stars and black holes is compatible with observa-tions. What are these parameters?We illustrate these parameters with an example of one of the tracks in the Scenario Machine that leads to the merging of two black holes (Figs 5 and 6) [12, 42].

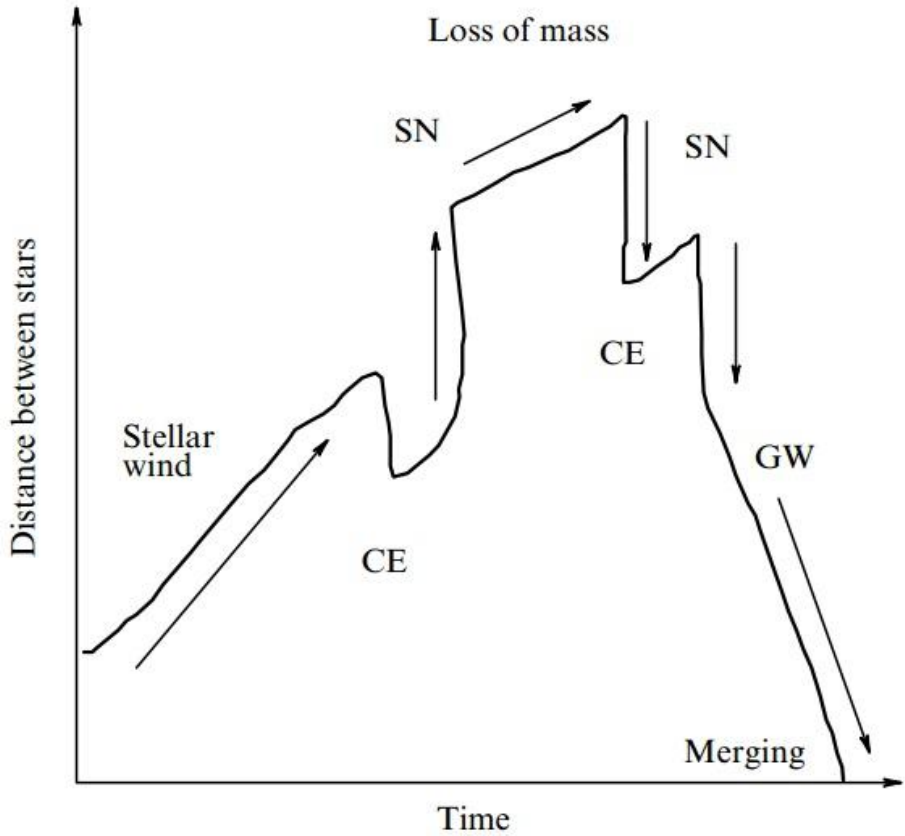

*Figure 5. Qualitative description of the varying distance between stars in a binary system. The stellar wind (loss of mass) increases the distance. However, an asymmetric supernova explosion (SN) and the common envelope stage (CE) occurring when one of stars is effectively absorbing the other make the stars approach each other (merge), giving rise to gravitational waves (GWs).*

Einstein's formula (1) for the gravitational wave luminosity says that the rate of orbital momentum loss is determined by the distance between components and their masses. It follows that processes changing the distance between stars play a key role in the final merging of relativistic stars.

We consider Figs 5 and 6. Two massive stars are originally formed as main-sequence stars. In our case, they are two blue stars with masses close to the upper mass limit for currently forming stars. Due to the high luminosity, the stars lose their mass in the form of stellar wind. In this case, the stellar wind not only blows away the mass of the star but also reduces its orbital moment. If the stellar wind is sufficiently fast (this is the case for massive blue stars, where the wind speed approaches 1500-2000 km $s^{-1}$) and spherically symmetric, then a binary system begins to `dissolve'. Qualitatively, it can be explained as follows. Given that the outflow is isotropic, the star loses its mass faster than its orbital momentum and hence the specific momentum increases, which is possible only if the stars are moving apart.

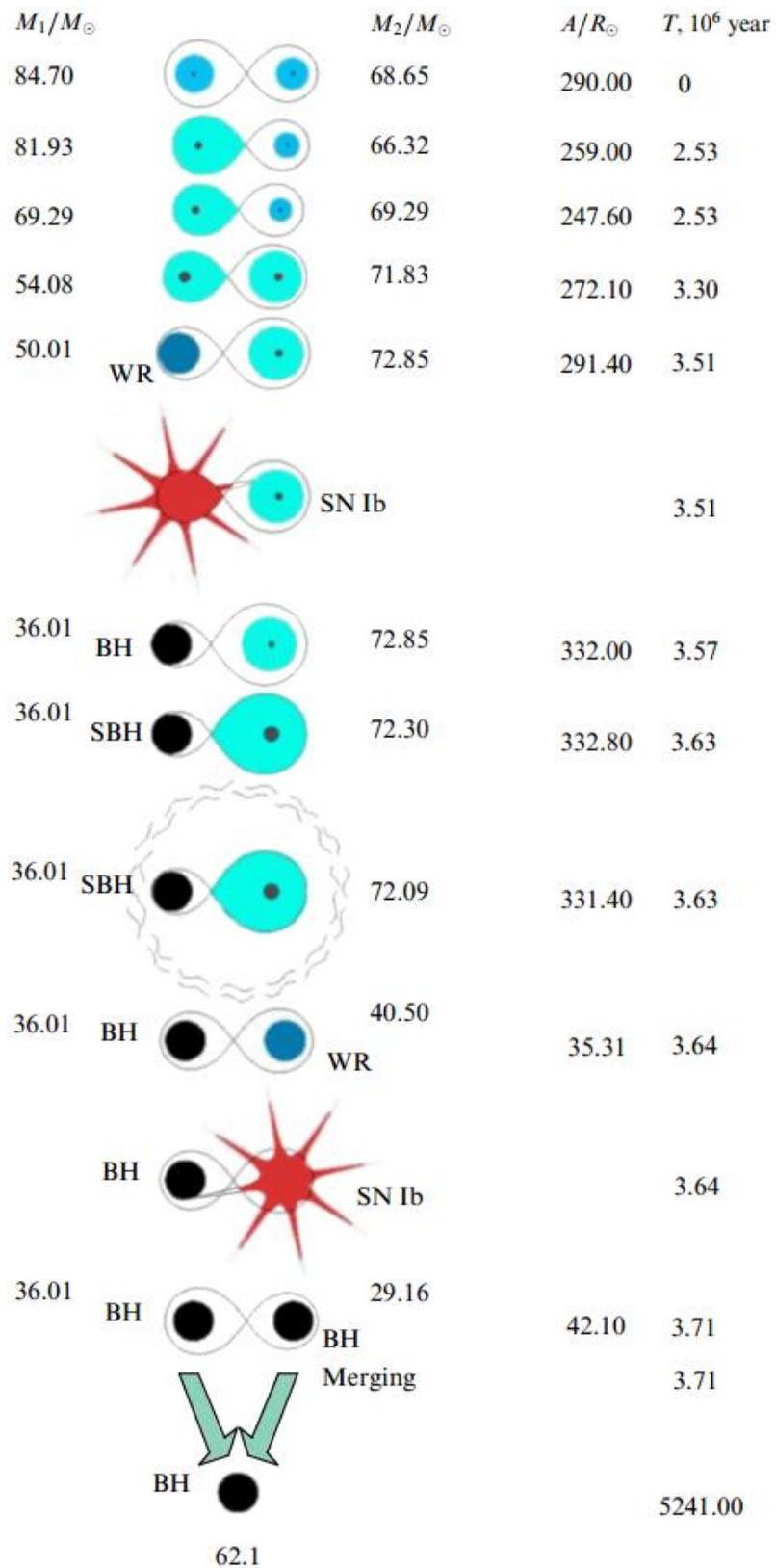

*Figure 6. One of possible evolutionary tracks calculated using the on-line version of the Scenario Machine [43]. Two stars in the main sequence with masses $M_1$ and $M_2$ begin to evolve; T is the beginning time of each stage, A is the distance between the stars, R is the solar radius.*

Thus, it turns out that the stellar wind is the enemy of the merging because it increases the distance between stars while reducing their mass, which is also included in the Einstein formula. Unfortunately, the self-consistent theory of stellar wind has not been elaborated yet. Instead, we use various formulas based on conservation laws quantitatively con-firmed by observations. For example, a very popular formula in the 20th century stated that the momentum of matter blown away by the stellar wind is proportional to the photon momentum emitted by the star [43]:

$$\frac{dM_w}{dt} V_w \sim \frac{L}{c},$$

where $dM_w/dt$ is the rate of the mass loss in the form of stellar wind, $V_w$ is the stellar wind velocity, and $L$ is the luminosity.

In the last formula, the stellar wind rate is proportional to the luminosity divided by the stellar wind velocity, which is approximately the escape velocity of the star. Equating the wind velocity with the tripled escape velocity near the star surface, we find that this formula is well confirmed by observations of hot stars in our Galaxy [44]. It directly follows from this formula that massive stars do not have enough time to lose a considerable part of their mass (< 10%), and therefore cannot be drawn sufficiently far apart. After the initially more massive star leaves the main sequence and fills the Roche lobe, catastrophic events in a binary system begin to occur.

First of all, if the component mass ratio is large enough, the matter from the first component cannot fall onto the second component because of thermal inertia. It is more likely to result in the formation of a so-called common envelope and a sharp increase in dynamical friction in the binary system such that the components begin to approach each other. But this does not happen in our case where the masses are roughly equal. Then, after releasing the envelope, only the helium core remains in place of the first star (all the hydrogen fuel runs out and the star is not a supergiant anymore).

Helium stars, called Wolf-Rayet (WR) stars, are observed by astronomers. The lifetime of WR stars is one order less than that of hydrogen stars. Therefore, in a few hundred thousand years, when the carbon-oxygen shell is formed, the first star begins to collapse, whereas the envelope explodes as a supernova. At that moment, the binary system fate is exposed to great danger, because an instantaneous loss of more than half of the mass destroys it. Hence, there is another very important and ill-defined parameter $k_{BH}$ - the portion of matter falling onto the black hole during the supernova stage (it is this parameter that varies along the horizontal axis in Fig. 5). This is a parameter in our problem.

There is also another parameter. The explosion can be asymmetric, such that the assumption on the loss of half the mass cannot be true anymore, and the problem becomes even more ambiguous. After the black hole is formed, there is a system of a blue supergiant star and a black hole! There is such a system in our Galaxy: the black hole candidate number one (chronologically) Cygnus X-1.

The evolution of the second star goes along the same lines. However, in this case, filling the Roche lobe by the giant and the rapid flow from the star onto the much less massive black hole yield the common envelope. This is followed by a second explosion and the formation of a binary black hole. As we can see, the change in the distance between stars in a binary system is a competition between two factors: the mass loss makes the components recede from each other, while the formation of the common envelope makes them approach each other, leading the black hole system to merge.

Hence, the basic parameters of the system are the distribution function of the components [6] with respect to the mass ratio $\varphi(q)$ ($q = M_2 / M_1 < 1$), the effectiveness of the common envelope, and the stellar wind power. There are also better-defined parameters, such as the distribution function of the initial Salpeter mass and the distribution of binary systems on the main sequence along the axes, which are considered to be known much better [45].

It seems that with so many unfixed parameters, it is absolutely impossible to find the black hole merging rate in our and other galaxies, even approximately. Notably, this explains why some authors claim that black holes do not merge at all.

However, the main idea behind the Scenario Machine is that we tried not just to calculate presumable properties and statistics of unobservable evolution stages of binary stars (note that black hole merging was not observed until 14 September 2015!), but also to correctly describe the observable stages. The more observable stages we use as frame points, the more exact our predictions of the proper-ties of unobservable stages are. For example, there is a black hole with a blue supergiant in our Galaxy. It follows that in modeling the Galaxy we have to ensure that our artificial sky has at least one black hole candidate with a blue supergiant.[4]

In our Galaxy, moreover, there are a few dozen X-ray pulsars - accretive neutron stars with massive blue star companions - and about 10 radio pulsars with neutron star companions. All of them are at different stages of the same evolution, which is the massive binary system evolution. On the other hand, if we suppose that one of blue stars generated a neutron star rather than a black hole, then we have to observe binary radio pulsars with black hole companions. However, they have not been detected yet. This fact should be taken into account when considering the population synthesis.

Of course, such a complicated modeling required a huge amount of astrophysical research, programming work, and calculations. But by 1997, when we finished the black hole calculations, we had 15 years of experience on population synthesis. We were therefore able to develop a program pack that allowed us to calculate neutron stars, black holes, and usual stars. We stress that the Scenario Machine has no analogues worldwide able to analyze the population synthesis of binary stars at the same level of detail. In particular, other programming packs do not take the rotational evolution of magnetized neutron stars into account. Sometimes, when comparing predictions of the Scenario Machine and other population synthesis codes, this gives rise to totally different merging rates of relativistic stars.

In [2], to obtain the most reliable prediction of the first events on gravitational wave interferometers, we calculated a weak stellar wind scenario by varying all poorly known parameters listed above. Particular attention was paid to the fact that our Galaxy should have at least one binary

---

[4] Of course, there may be doubts whether having just one black hole can be considered statistical. But the point is that systems consisting of a blue supergiant and a black hole candidate have been detected in nearby galaxies with a similar rate of star formation.

In [2], to obtain the most reliable prediction of the first events on gravitational wave interferometers, we calculated a weak stellar wind scenario by varying all poorly known parameters listed above. Particular attention was paid to the fact that our Galaxy should have at least one binary system of the Cygnus X-1 type and no pulsars with a black hole companion for 1000 single radio pulsars [31, 46]. We recall that such systems have not yet been discovered, even though about 2000 single radio pulsars have been detected.

It is obvious that the first condition imposes a lower bound on the black hole merging rate, while the second places an upper bound on it. The large grey region in Fig. 1 resulted from calculations using all the listed parameters. In the weak stellar wind scenario, we therefore have a definitive result: first events on LIGO-Virgo-type detectors should involve black holes! In [3, 4], using the scenario with a big mass loss in the form of stellar wind, we arrived at the same result.

What is a strong stellar wind? As we have already noted, the stellar wind power is crucial for the fate of a binary system and the probable merging of its evolution products. In the late 20th century, the concept of a strong stellar wind was introduced in accordance with the so-called energy formula

$$\frac{1}{2}\frac{dM_w}{dt} V_w^2 = L.$$

Obviously, in this case, the stellar wind increases by $c/V_w$ times. According to this formula, a star with a mass of the order of 100 $M_\odot$ loses more than 90% of its mass and can produce just a neutron star rather than a black hole of a mass (20-30) $M_\odot$. Of course, with this stellar wind, the binary system components move away from each other as early as during the main-sequence stage (hydrogen burning). It seems that they have no chance to merge. However, the beauty of the Scenario Machine is that with such a wind, if no additional constraints are imposed, no black holes would merge and, moreover, the sky would not contain Cygnus X-1 type objects or, for example, the Hulse-Taylor pulsar, for the discovery of which the Nobel prize has already been awarded.

In [4], we considered the strong-wind scenario and showed that if we want to obtain data corresponding to the observable picture of relativistic binary stars, then an anisotropy of a collapse during the supernova explosion has to be introduced. It turns out that small kick velocities (similar to those of a gun firing) of the order of 150- 200 km s$^{-1}$ destroy binary systems of neutron stars and bind binary black holes.

The Nobel prize winner H.Bethe appreciated this effect. At the end of his fruitful career, he studied population synthesis to find the merging rate of relativistic stars in the Universe. In one of his last papers, Bethe wrote [41]: ``In particular, these authors (Lipunov, Postnov, Prokhorov, 1997 [4]) find that introducing kick velocities increases their merging rate by an order of magnitude. Portegies Zwart, Yungelson [47] find zero merges without kick velocities."

However, the paradox is that binary neutron stars in the Universe merge much more often than black holes. For example, the merging rate of neutron stars in a Milky Way type galaxy is 1 event every 10,000 years (up to a factor of 2-3 [20]), while the black hole merging rate is much less: one every few hundred thousand years. However, black hole mergings are detected more frequently (see Fig. 1). The fact is that a detected signal - a displacement of the interferometer arms proportional to the gravitational wave amplitude h (dimensionless number) - is given by contracting and stretching harmonics of a known form. Given that the signal form is known a priori, we increase the sensitivity by many times. But in this case, the signal magnitude is proportional to the so-called chirp mass [5]

$$\mathcal{M} = \frac{(m_1 m_2)^{3/5}}{(m_1 + m_2)^{1/5}}$$

raised to the power 5/6. Because the amplitude decreases as the inverse distance to the source, the detected volume for black holes is proportional to the signal power cubed, which turns out to be a few thousand times bigger for black holes. This is why the `Loch Ness monster' (see Fig. 1) is above the merging neutron stars.

## 6. Why did the black hole masses turn out to be much higher than expected?

The anomalously high (as many believe) black-hole masses were extensively discussed immediately after the discovery of the first merging black holes; the mass of each turned out to be equal to 30$M_\odot$. Indeed, the statistics of so-called black hole candidates discovered during the last 30 years shows that the average mass of a black hole in binary systems is of the order of (6-7)$M_\odot$ [48]. However, we note that most of the candidates with a relatively well established mass are so-called X-ray novae - binary systems, whose optical components are dwarf stars with masses of the order of or less than the solar mass. However, such systems do not produce binary black holes and are not direct ancestors of LIGO detector events. As we have emphasized, binary black holes are produced from massive stars able to generate black holes on their own, with regard to which it seems that we need to explain why black hole masses in X-ray novae systems are relatively small rather than why the GW150914 black holes are big. We return to this subject below. We now discuss other factors contributing to the large mass of the first detected merging black holes.

What is the meaning of the big mass of the gravitational wave burst event GW150914? The first obvious conclusion is that this event is the result of the massive binary system evolution model with a weak stellar wind considered in our paper [2]. Figures 5 and 6 show a possible evolutionary track leading to merging black holes with masses 29$M_\odot$ and 36$M_\odot$. The track is generated by the on-line version of the Scenario Machine [43].

As we have noted, the resulting black holes are massive enough and match observations, because the stellar wind is (relatively) weak. In principle, before the mid-1990s, most authors preferred to consider the de Jager stellar wind model [44], which is essentially based on observation data.

A strong stellar wind was introduced into the evolution theory of extremely massive stars with initial masses of more than (40-50) $M_\odot$ by Woosley [49]. We stress that the energy rather than momentum stellar wind was introduced `by hand'. It was neither self-consistently calculated nor observationally confirmed. This question remains debatable and most probably will be solved in favor of a relatively weak stellar wind.

---

[5] The English language literature uses this term `chirp mass' originating from the popular analogy between the gravitational wave burst and the chirping Universe. For example, the gravitational wave frequency in black hole merging is low enough and only at the very end may reach several hundred hertz.

On the other hand, it is known that low-metallicity stars have weaker stellar winds; the radiative pressure is propor-tional to the cross section of the interaction of photons with atoms and ions of matter. The cross section sharply increases even when there are minor amounts of metal in the stellar atmosphere. There are semiphenomenological formulas describing the dependence of the mass loss rate on the metallicity of the stellar atmosphere. Such weakly metallic stars should be born first (third generation) in our Universe or in dwarf galaxies with reduced metallicity like Magellanic Clouds.

The evolution scenario of the third-generation binary stars [50, 51] was not considered in our paper [2]. However, our calculations in [2] can be directly used in this case, because the stellar wind blows away just a small amount of the progenitor mass, which is typical for stars with low heavy-metal content.

Moreover, there is a selection effect giving rise to a high probability of observing events with an anomalously large total mass of the black holes [12]. The density of events with an amplitude h can be found by considering a spherical shell of radius r. It is obvious that $dN(r|h_0) = 4\pi r^2 dN(h_0)\, dr$, where $dN(h_0)$ is the number of coalescence events per unit volume with the gravitational amplitude $h_0$ per unit distance. Passing to the observable amplitude $h = h_0/r$, we find that

$$dN(h|h_0) = 4\pi\, dN(h_0)\, \frac{h_0^3}{h^4}\, dh.$$

The resulting relative probability distribution of gravitational amplitudes can be found by integrating over all $h_0 = \Gamma M^{5/6}$ (where $\Gamma$ is a factor depending on the distance to the binary system and the gravitational wave frequency) or over all chirp masses:

$$dN(h) = \frac{4\pi}{h^4}\int dN(h_0)\, h_0^3\, dh_0 = \frac{4\pi}{h^4}\frac{5}{3}\Gamma^4 \int dN(M)\, M^{7/3}\, dM.$$

The probability of registering an event with the amplitude larger than some threshold value P is given by

$$P(h > \Pi) \approx \frac{20\pi}{\Pi^3}\Gamma^4 \int dN(M)\, M^{7/3}\, dM.$$

It is obvious that an additional mass $M^{7/3}$ greatly increases the probability of observing events related to a big chirp mass of merging relativistic objects.

Shifting the median of the expected distribution towards bigger masses with a total mass of more than $50 M_\odot$ seems quite normal.

As an illustration, Fig. 6 shows one possible track generated by the on-line version of the Scenario Machine. It is based on the weak stellar wind scenario. We suppose that this scenario can be applied both to first-generation stars for which the stellar wind can be anomalously weak and to massive stars that are currently forming. We see that the system goes through two supernovae bursts and the common-envelope stage in about 3.7 million years. However, the merging occurs only in 5 billion years.

There is one more important fact supported by the GW150914 event parameters: the proximity of the masses of two merging black holes. This implies that the initial mass ratio of massive binary stars was also close to unity. More-over, this is confirmed by the track. This is a nice argument in favor of the massive binary system distribution function of the mass ratio $q=M_2/M_1 < 1$ with a maximum at unity. The function $\varphi(q) \sim q^2$ was proposed in [53] and used in the Scenario Machine as the preferred one.

We now return to the question of why the black hole masses in low-massive binary systems are so small. The fact is that for a small initial mass ratio $q = M_2/M_1 <\sim 100$, a dwarf star does not have time to form (to arrive to the main sequence). Instead, it is `evaporated' by the blue giant with a luminosity a few million times greater than that of the dwarf star. Indeed, the protostar concentration stage lasts for the thermal time:
$t_{th} \sim 3 \cdot 10^7 (M_2/M_\odot)^2$.

At this stage, the star radius is determined by the complete absence of ionization, as a mirror counterpart of how the recombination makes the Universe transparent. To ionize all hydrogen atoms, the energy $13.6\, M_2/m_p\, [eV]$ is required (where $m_p$ is the proton mass). On the other hand, the gravitational energy of the star is $GM_2^2=R_2$. Equating the two values, we find the protostar radius $R_2 \sim 150 R_\odot (M_2/M_\odot)$. We now calculate the energy emitted by the massive blue star and captured by the low-massive protostar. The optically opaque protostar captures the energy $L_2=(1/4)(R_1/a)^2 L_1$, where a is the distance between stars. The stellar wind arises after absorption and reheating. It evaporates the protostar with the minimal rate determined by the momentum conservation law: $dM_2/dt \sim (L_2/3)\, v_p c$ [44], where $v_p$ is the parabolic velocity. The total mass loss is proportional to the massive blue star lifetime $T_1$. Assuming that the total mass loss is equal to the dwarf mass $M_2$, we find that the low-mass protostar evaporates if $a < 450 R_\odot (M_2/M_\odot)$. It follows that not all these systems survive; hence, they do not produce X-ray novae. Thus, the only condition for a dwarf star to survive in a binary system with a blue supergiant is the following: the nuclear time of a massive star should exceed the thermal time of the dwarf protostar, i.e.

$$q_0 = \frac{M_2}{M_1} \gtrsim \frac{1}{17}.$$

Using the mass of the smaller component $M_2 <\sim M_\odot$, we find that there are no massive blue progenitors with masses exceeding $(17-20)M_\odot$ among X-ray novae. We emphasize that this is a mass of the progenitor in the main sequence. It follows that the black hole mass is half this value, i.e., it coincides with the average mass of X-ray novae.

Thus, the small average black hole mass observed before is related to the fact that the massive progenitors evaporate their companions, thereby destroying the binary systems. It follows that they fall out of the statistics. However, the relatively large mass of the event GW150914 well fits the computational data of the massive star evolution scenario with a weak stellar wind and with the selection effects induced by increasing the merging detection horizon with the growth of the total mass of a binary system properly taken into account.

**7. Gamma-ray event detected by the Fermi observatory**

The EM follow-up program for the gravitational wave LIGO experiment includes all X-ray and gamma-ray observatories, in particular, the American±Russian experiment Konus-Wind (Interplanetary Network), INTEGRAL (Interna-tional Gamma-ray Astrophysics Laboratory), Swift-BAT (Burst Alert Telescope), Swift-XRT (X-ray Telescope), the Fermi gamma-ray observatory, and the Japanese experiment MAXI (Monitor of All-sky X-ray Image) [1]. However, only the Fermi observatory reported a short (less than 1 s) weak gamma-ray burst 0.4 s after the gravitational wave trigger on the GBM (Gamma Burst Monitor) detector [39]. A burst with an energy of $\sim 3 \cdot 10^{-7}$ erg was later identified in the archived recordings of the gamma-ray background after the G184098 alert was received.

The Fermi event localization region is shown in Fig. 4. The common region of the Fermi and LIGO intersecting error square is 90% covered by the MASTER-SAAO telescope (SAR) observations. This region was detected by the MASTER telescope only. On the other hand, we did not detect any optical radiation brighter than 19 magnitude, which can possibly be related to the gravitational wave event GW150914/G184098 [12]. We do not discuss here how real the Fermi event is.

Instead, we discuss the possible relation between the gamma-ray burst and the black-hole merging. We have noted that the radiation of standard gamma-ray bursts is strongly anisotropic, and the probability of gravitational and gamma-ray bursts being detected simultaneously is much less than 1/100. Moreover, the gamma-ray burst energy as seen from the distance to the GW150914 event is $E_{Fermi} \sim 2 \cdot 10^{49}$ $erg$ $s^{-1}$, which is much less than typical values for the isotropic energy of gamma-ray bursts. Thus, this hypothesis, actively discussed in [54], is to be rejected.

According to general relativity, the electromagnetic radiation arising from the merging of two noncharged black holes can result only from the presence of additional matter in binary black holes or in their neighborhood. For example, in 1984, Lipunov and Sazhin noted in [55] that a powerful electromagnetic burst could result from the merging of supermassive black holes surrounded by a dense star cluster, which is present in almost all galactic nuclei. Obviously, this is not the case with GW150914/G184098.

However, some matter surrounding black holes can be accumulated as a result of interstellar gas accretion at the stage preceding the merging. Assuming that the typical coefficient of energy emission near accretive black holes is 10% [54], we find that the required mass is of the order of $\Delta M \sim 10^{-3}$ $M_\odot$, which is close to the mass of Jupiter. It seems that this mass is quite small, but with the time lag $\Delta t \sim 0.4$ s corresponding to the distance $10^{10}$ cm taken into account, we find that the plasma density near black holes must be close to the density of water: $\rho \sim \Delta M / (c\Delta t)^3 \sim 1$ g cm$^{-3}$. In fact, this is Jupiter's density! However, it is hard to imagine that such a ring or planet is present in the system of two blue supergiants. Part of the matter could be captured when the typical distance between black holes is much less than $c\Delta t \sim 10^{10}$ cm. Due to continuous gravitational radiation, the duration of this stage cannot exceed

$$t \sim \frac{I\Omega^2}{2L} \sim \left(\frac{A}{10^{10}\ \text{cm}}\right)^4 \Big/ \left(\frac{M}{60 M_\odot}\right)^3 \times 1\ \text{year}.$$

The maximum mass that can be accumulated in a year is $\Delta M \sim (dM/dt) \times 1$ year, while the accretion rate can be estimated by the Bondi-Hoyl formula [42]:

$$\frac{dM}{dt} = \frac{\pi(2GM)^2}{v^3}\rho \sim 10^{-12} M_\odot\ \text{year}^{-1} \times \left(\frac{M}{60 M_\odot}\right)^2 \frac{\rho}{10^{-24}\ \text{g cm}^{-3}} \left(\frac{V}{10\ \text{km s}^{-1}}\right)^{-3}$$

where $M$ is the total mass of black holes, $V$ is the velocity of black hole motion relative to the interstellar medium in the host galaxy, and $\rho$ is the interstellar density.

It is clear that one year is insufficient to accumulate a mass of $10^{-3} M_\odot$. Thus, it must be recognized that the event detected by the Fermi observatory is apparently unrelated to the LIGO GW150914 event.

## 8. Conclusion

The detection of the gravitational wave information channel is the breakthrough that takes humankind across a frontier into a new era. It can be compared only with Galileo's discovery when he turned his telescope toward the night sky. Undoubtedly, gravitational wave astronomy will become a statistically valid science in the nearest future. It will be possible to investigate in depth the most powerful processes associated with relativistic stars merging in our Universe. However, there are more fundamental problems. Back in the 1970s, Grishchuk [56] showed that the Universe is full of relic gravitational waves generated at the Universe's birth. Their registration would help to understand how our space-time was created. Generally speaking, modern and future interferometers are designed to detect the cosmological back-ground. But the question arises as to whether cosmological gravitational waves fade on the background of the `modern' radiation generated in our and other galaxies.

About half of all stars in the Universe are binary. The Universe is full of gravitational waves, while Earth literally swims in this gravitational sea. In 1965, the Soviet astronomer Mironovsky [57] first tried to detect which gravitational frequencies stir up the sea. It tuned out that the maximum amplitude is generated by the most narrow normal stars of the Ursa Major W type. With a period of a few hours, these stars come so close to each other that their surfaces are in contact and generate gravitational waves with the dimensionless amplitude $h \sim 10^{-20}$. It follows that binary stars can pose problems for the detection of the cosmic background.

There was a task to calculate the complete radiation spectrum of all binary stars in the Universe. The calculation was done in 1986 by Soviet astrophysicists [58]. It turned out that the leading contribution comes from binary stars of our Galaxy. Moreover, they suppress the cosmological back-ground in a wide frequency range from $10^{-5}$ Hz to several Hz. However, our Galaxy is flat; hence, its images in the gravitational and electromagnetic skies are about the same, having the form of a specific Milky Way gravitational wave.

Apart from the galactic plane, the main signal comes from binary stars of distant galaxies distributed over the sky quite homogeneously. Nevertheless, there are `windows' at the edges of the spectrum through which the relic background can be seen. This gives hope that someday we will know how our Universe was born.

In order to detect these low-frequency waves, interferom-eters with giant mounts are necessary, which must be built in outer space. The LISA project (Laser Interferometer Space Antenna) is expected to launch several spacecraft that will form a giant laser interferometer in the Solar System with an arm several million kilometers long. Perhaps we will then find out how our Universe began.

**Acknowledgments**
The author thanks Kip Thorne for the discussions. Also, the author speaks well of Leonid Petrovich Grishchuk, who predicted the cosmological gravitational wave background (see [56]). Unfortunately, he did not live a few years longer to see the discovery of gravitational waves. It is thanks to these persons that we performed the computations described in this paper. This work was supported by the RFBR grant 150207875.